\documentclass[conference,10pt]{IEEEtran}

\usepackage[T1]{fontenc}
\usepackage{gensymb}
\usepackage{todonotes}
\usepackage{booktabs}
\usepackage{multirow}
\usepackage{amsmath}
\usepackage{afterpage}
\usepackage[noadjust]{cite}
\usepackage{subcaption}
\usepackage{enumitem}
\makeatletter
\g@addto@macro\normalsize{%
  \setlength\abovedisplayskip{4pt}
  \setlength\belowdisplayskip{4pt}
}
\makeatother

\makeatletter
\if@twocolumn%
\else
\fi
\makeatother

\usepackage[cmintegrals]{newtxmath}
\usepackage{amsmath}
\usepackage{algorithm}
\usepackage{algpseudocode}
\usepackage{microtype}

\usepackage{bm}

\usepackage{url}
\usepackage{bbm}
\usepackage{graphicx}
\newcommand{\subparagraph}{}

\begin{document}
\title{Resource Sharing Among mmWave Cellular Service Providers in a Vertically Differentiated Duopoly}
\author{\IEEEauthorblockN{Fraida~Fund, Shahram~Shahsavari, Shivendra~S.~Panwar, Elza~Erkip, Sundeep~Rangan}
\IEEEauthorblockA{Department of Electrical and Computer Engineering,
NYU Tandon School of Engineering 
\\ \{\url{ffund, shahram.shahsavari, panwar, elza, srangan}\}@nyu.edu
}
}

\maketitle

\begin{abstract}

With the increasing interest in the use of millimeter wave bands
for 5G cellular systems comes renewed interest in resource
sharing.
Properties of millimeter wave bands
such as massive bandwidth, highly directional
antennas, high penetration loss, and susceptibility to
shadowing, suggest
technical advantages to spectrum and infrastructure sharing
in millimeter wave cellular networks.
However, technical advantages do
not necessarily translate to increased profit
for service providers, or increased consumer surplus.
In this paper, detailed network simulations are used to
better understand
the economic implications of resource sharing
in a vertically differentiated duopoly market for
cellular service.
The results suggest that resource sharing is less often
profitable for millimeter wave service providers compared
to microwave cellular service providers, and
does not necessarily increase consumer surplus.

\end{abstract}

\IEEEpeerreviewmaketitle

\section{Introduction}
\label{sec:introduction}

With the introduction of millimeter wave (mmWave) cellular systems
as a candidate for the next generation of mobile networks,
there has been renewed interest in problems related
to sharing resources, i.e. spectrum and base stations (BSs).
Early results suggest that
compared to conventional microwave frequencies,
the massive bandwidth,
increased spatial degrees of freedom, and greater path loss
increase the technical benefits of resource sharing in mmWave frequencies
\cite{andrews-sharing,matia,european-sharing,andrews-sharing-second}.
Some have suggested that these technical gains will
translate to economic gains. For example,~\cite{andrews-sharing}
claims that it is economical for mmWave service providers to share resources
because they can offer the same quality of service while
licensing less spectrum. Similarly,~\cite{andrews-sharing-second} considers a scenario
where a primary mmWave spectrum holder can earn additional revenue by
licensing the spectrum in a secondary market with the condition of
restricted interference to its own users.
However, even if service providers can reduce licensing costs or earn revenue
from secondary licensing while keeping quality of service the same,
resource sharing can affect profits
if it shifts demand to a competing service provider,
or if it changes the market dynamics in a way
that forces down the price.
None of~\cite{andrews-sharing,matia,european-sharing,andrews-sharing-second}
consider the effects of demand and competition.

Some of the literature in both engineering
and economics disciplines addresses
economic and regulatory aspects of resource
sharing in conventional cellular
networks~\cite{dublin2,coalitionsharing,Markendahl2012opetition,Meddour:2011:RIS:1975024.1975472,jsacCoop}.
However, the fundamental technical differences between
mmWave and microwave frequencies
also affect the markets for these services.
For example, an early economic perspective on mmWave
networks~\cite{Nikolikj:2015:SBP:2822170.2822198}
suggests that their limited coverage range
is a key challenge for their cost efficiency.
Similarly, our previous work~\cite{funds3} shows that
the dynamics of demand and ease of market entry,
and the effect of unlicensed spectrum or open association
small cells on these, are different
in mmWave networks (compared to microwave small cell networks).
Thus, we posit that the economic impact of resource sharing
may also be different.
To gain a fuller understanding of the benefits of resource
sharing in mmWave networks,
we need to identify the specific impact on
quality of service, and then understand how this
affects the demand, price, and cost of service.

The goal of this work, therefore, is
to understand the strategic decisions of
wireless service providers with respect to resource sharing
in mmWave 5G cellular networks.
We apply the economic model of \emph{compatibility
of network goods}~\cite{katz1985network,economides1997compatibility} to
resource sharing in mmWave cellular networks.
We describe a duopoly game involving two vertically differentiated
cellular service providers,
and compare mmWave and microwave networks
with respect to service provider profits, market coverage,
consumer surplus, and price of service with
and without resource sharing.

The rest of this paper is organized as follows.
We begin with a brief introduction to the economic
framework used in this paper, in Section~\ref{sec:economic}.
In Section~\ref{sec:system}, we describe the
wireless system model,
and show simulation results in Section~\ref{sec:simulation}.
In Section~\ref{sec:duopoly},
we describe a duopoly game involving two vertically differentiated
network service
providers, and compare the benefits of resource
sharing decisions to consumers and service providers
in mmWave and microwave networks.
Finally, in Section~\ref{sec:conclusion}, we conclude with a discussion
of the implications of this work and areas of further research.

\section{Economic foundations}
\label{sec:economic}

In this paper, we use an economic model that includes several
important considerations for resource sharing in cellular networks:
\begin{itemize}
\item \textbf{Network effect}. This captures the
effect of a large network service provider (NSP) with more
subscribers typically having more base stations and spectrum than
a competitor with a smaller market share.
\item \textbf{Compatibility}. This models the decision
of an NSP to share resources
(base stations and spectrum) and potentially increase the value of
its service to subscribers, or to preserve its own market
power by not sharing resources.
\item \textbf{Vertical differentiation}. This models
the ability of an NSP to distinguish itself
from competitors in aspects of its service other
than network size and price.
\end{itemize}
The network size and the extent of the network effect, as
well as the compatibility (resource sharing) decision, 
determine the consumer's expected data rate, 
and the vertical differentiation captures other
factors affecting consumers' decisions such as customer service and
availability of desirable handsets.

In economics, a network good or service~\cite{economides1996economics}
is a product for which
the utility that a consumer gains from the product
varies with the number of other consumers of the product
(the \emph{size of the network}).
This effect on utility - which is called the
\emph{network externality} or the \emph{network effect} -
may be direct or indirect.
For example, consumers of a telephone network, which is more
valuable when the service
has more subscribers, benefit from a direct network effect.
The classic example of an indirect effect
is the hardware-software model,
e.g. a consumer who purchases an Android smartphone
will benefit if other{} consumers also purchase
Android smartphones, because this will incentivize
the development of new and varied applications for
the Android platform. The network externality
may also be negative, for example, if an Internet
service provider becomes oversubscribed, its subscribers
will suffer from the congestion externality.
Network effects and the standard models
for understanding them have been empirically
validated in a variety of industries,
including the markets for
mobile phones~\cite{madden2004dynamic},
fixed broadband~\cite{lee2011cross},
personal digital assistants~\cite{Nair2004},
DVDs~\cite{dranove2003dvd},
home video games~\cite{shankar2003network},
ATMs~\cite{NBERw4048},
and fax machines~\cite{economides1995critical}.

Consider a set of consumers in a market for a good.
Total demand for the good, $n$, is normalized
so that $n=1$ when all consumers purchase
the good, and $n=0$ when no consumers purchase
the good. Heterogeneous consumers are parameterized
by the ``taste parameter''
$\omega$, $0 \leq \omega \leq \hat{\omega}$.
A consumer with
a high $\omega$ is willing to pay more for a
high-quality good.

In the case of a non-network good,
the utility of a consumer of type $\omega$
may be modeled as $u(\omega, p) = \omega - p$,
and the consumer is indifferent between purchasing the good
or not when price $p = \omega$.
For $p > \hat{\omega}$, no consumer will purchase the
good ($n=0$), because it is too expensive even for consumers
with the highest value of $\omega$. At $p=0$,
all of the consumers will purchase the good ($n=1$).
The demand curve, which indicates what portion
of the consumers will purchase the good at a given
price, has a negative slope for most
non-network goods,
because the quantity demanded $n$
increases as the price of the good $p$
decreases.
Conversely, to increase demand for a typical
non-network good, the producer must reduce its price.

In the case of a network good,
a consumer of type $\omega$ purchasing the good at price $p$
gains utility $u(\omega, n, p) = \omega h(n) -p$,
where $n$ is the network size,
and $h(n)$ is a network externalities
function indicating how consumer utility
scales with $n$. For a positive network externality,
$h(n)$ increases with $n$, and for a negative network
externality $h(n)$ decreases with $n$.
With a positive network externality, the demand curve
may have a positive slope. Unlike a non-network good,
where increasing the price reduces the demand
for the good, the value of a network good
varies with the number of consumers, so as more
units are sold, the value increases and the producer
can raise the price.
For a more detailed overview,
see~\cite{economides1996economics}.

Producers of a network good with a positive network 
externality therefore prefer to
increase the size of their network.
In addition to selling more units of the
good, producers may increase network size by making
\emph{compatible} goods~\cite{katz1985network,economides1997compatibility}.
When two network goods are compatible, then the total network
effect for a consumer of either good is based on
the sum network size of both goods.
In other words, the network externalities
function for good $i$ is evaluated using the total network size
for all the goods:
$h( \sum_{j \in I} n_{j})$, where{}
$I$ is a set of firms producing compatible goods and $i\in I$.
Then the utility of a consumer of type $\omega$ purchasing
good $i$ at price $p$ is
$u(\omega, \sum_{j \in I} n_{j}, p) = \omega h( \sum_{j \in I} n_{j}) -p$, $i\in I$.
A producer of a network good with a positive network 
externality has incentives for and
against compatibility:
\begin{itemize}
\item \textbf{Network effect}: A compatible product is more 
valuable to consumers, since the argument to $h(\cdot)$ is greater.
\item \textbf{Market power}: Making a product
compatible increases the value of a competitors' product
and reduces the perceived difference between competing products,
potentially increasing price competition and forcing prices down.
Price competition occurs when competing products are 
very similar, so consumers' 
purchasing decisions are made mainly on the basis 
of which is cheaper.
\end{itemize}
Compatibility is often used to model a firm's
choice to use a proprietary technical standard or a common industry standard.
For example, the developers of a word processing application
might choose to use a proprietary file format so that all consumers
who need to open these files must purchase their software,
or they might choose to use an open standard so that their
users can share the files produced with their software
with users of other word processors.

The effect of the latter issue, loss of market power,
is partially mitigated by vertical
differentiation. With vertical differentiation,
a producer may choose compatibility and still distinguish itself
from competitors and reduce price competition by improving
the quality of its good in other ways
(not by increasing $n$). For example, consumers of
Android-based smartphones benefit from the network effects
due to consumers of all Android-compatible smartphones.
However, a producer of Android phones can distinguish
itself by selling handsets with better
hardware specifications than its competitors'.
With vertical differentiation, different providers
capture different portions of the market (e.g., low end
vs. high end).

In this paper, we use a model
of consumer utility in a market for network
goods with vertical differentiation,
described in~\cite{baake2001vertical}.
The willingness to pay of a consumer of type $\omega$
for good $i$ is $\omega q_i + q_i h( \sum_{j \in I} n_{j})$, where
$I$ is a set of firms producing compatible goods, $i\in I$,
and $q_i$ is a scaling factor that represents aspects of the good's
quality that are \emph{not} a function
of the network size. Firms that produce compatible goods
can distinguish themselves from one another
by choosing different quality levels. However, a firm that
produces a higher-quality good also has higher
marginal costs; the marginal cost to the producer of
producing one unit of good $i$ is $q_i$, its total cost
is $c(q_i, n_i) = q_i n_i$, and its profits are then
$\pi_i(q_i, n_i, p_i) = n_i p_i - q_i n_i$.

Given this economic framework, we are interested in modeling
mmWave network service as a network good, to better understand
the conditions under which mmWave network service providers will want
to share resources, and whether or not it is desirable for a regulator to
enforce resource sharing.
To answer these questions, however,
we must characterize $h(n)$, the function that determines
the network effect. In the next section,
we describe the simulation from which we
empirically derive $h(n)$ for mmWave and microwave small cell
networks.

\section{System Model}
\label{sec:system}

We describe the system model of the mmWave
and microwave network simulations.
The results of this simulation will be used
to parameterize the economic model for resource sharing.
Our simulation captures the following key characteristics
of mmWave and small cell microwave networks:
\begin{itemize}
\item {\textbf{Channel model}: We use the empirically derived
line of sight (LOS), NLOS and outage
probabilistic channel models for mmWave links
from~\cite{mustafa-channel}. For the microwave links, we use the microcell
channel model described in~\cite{UMichannel}.}
\item {\textbf{Directional transmission}: We use the antenna pattern model
described in \cite{heath-mm}.
For mmWave frequencies we use model parameters representing
an 8x8 antenna array at the base station (BS) and a 4x4 antenna array
at the user equipment (UE).
For microwave frequencies, we use the ITU model for the
BS antenna~\cite{UMichannel}, and an omnidirectional UE antenna.}
\end{itemize}

We consider a system with two NSPs. NSP $i \in\{1,\dots,I\}$
has bandwidth $W_i$, BSs
distributed in the network area using a
homogeneous Poisson Point Process (hPPP) with intensity $\lambda^B_i$,
and UEs whose locations are modeled by an
independent hPPP with intensity $\lambda^U_i$.

\begin{figure}
    \centering
\includegraphics[width=0.4\textwidth]{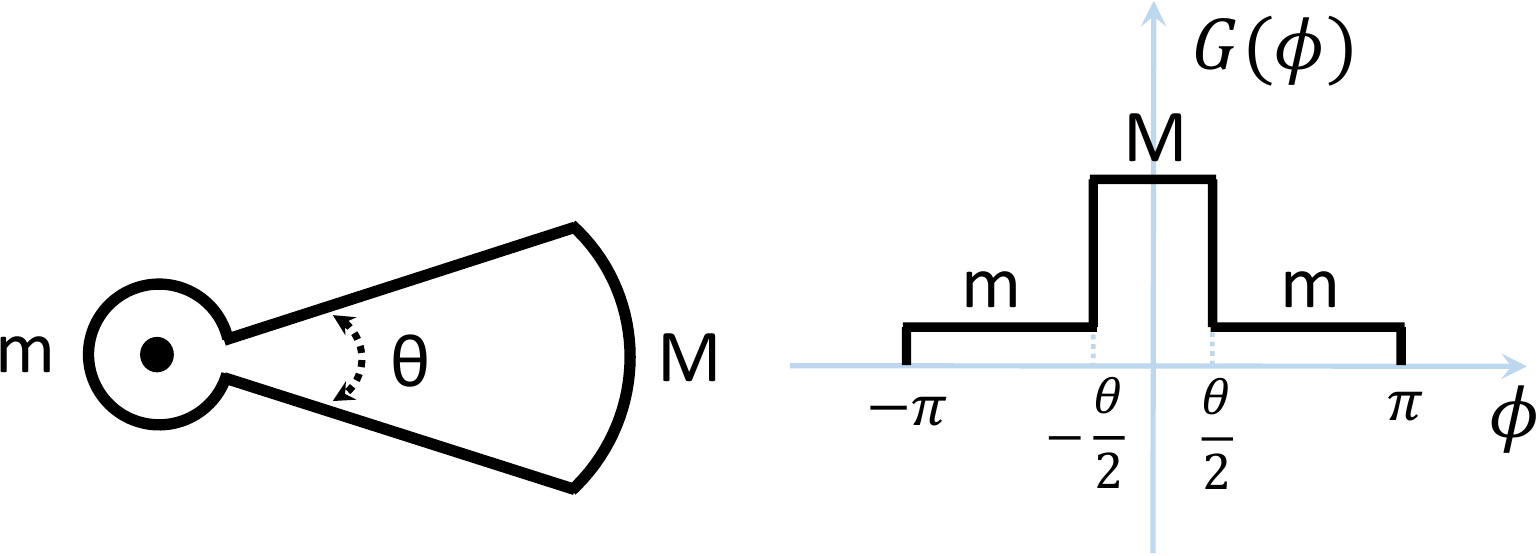}
\caption[]{Simplified antenna pattern with main lobe $M$, back lobe $m$ and beamwidth $\theta$.}
 \label{pattern}
\end{figure}

Both BSs and UEs may use antenna arrays for directional beamforming.
We approximate the actual array patterns
using a simplified pattern as in \cite{heath-mm,andrews-antenna}.
Let $G(\phi)$ denote the simplified
antenna directivity pattern depicted in Fig. \ref{pattern},
where $M$ is the main lobe power gain, $m$ is the back lobe
gain and $\theta$ is the beamwidth of the main lobe.
In general, $m$ and $M$ are proportional
to the number of antennas in the array and $M/m$ depends
on the type of the array. Furthermore,
$\theta$ is inversely proportional to the number of
antennas, i.e., the greater the number of antennas,
the more beam directionality. We let $G^B(\phi)$
(which is parameterized by $M^B$, $m^B$, and $\theta^B$) be
the antenna pattern of the BS, and $G^U(\phi)$
(which is parameterized by $M^U$, $m^U$, and $\theta^U$) be the antenna pattern
of the UE.

We model a time-slotted downlink of a cellular system.
For path loss, shadowing,
and outage, line of sight (LOS), and NLOS probability distributions,
we use models adopted from \cite{mustafa-channel} and \cite{UMichannel} for
mmWave and microwave links, respectively.
We assume Rayleigh block fading.
The data rate is modeled as
\begin{equation}
R=(1-\alpha)W\log_2 \Bigg(1+\beta \frac{PG^U(0)G^B(0)H}{N_fN_0W+I}\Bigg),
\label{rate-model}
\end{equation}
where $\alpha$ and $\beta$
are overhead and loss factors, respectively,
and are introduced to fit a specific physical layer to the
Shannon capacity curve \cite{gomezoptimal}.
Furthermore, $P$ is the BS transmitting power,
and $H$ is the channel power gain derived from the model
discussed above, incorporating the effects of fading, shadowing, outage, and path loss.
We assume perfect beam alignment between BS and UE within a cell,
therefore the antenna power gain (link directionality) is
$G^U(0)G^B(0)=M^UM^B$.
Finally, $N_f$, $N_0$, $W$ and $I$ are UE noise figure,
noise power spectral density, bandwidth, and interference power, respectively.

In the mmWave scenario,
each cell belonging to a given NSP $i$ reuses the whole bandwidth $W_i$
available to that NSP, with no coordination between
cells. Although it is possible to have strong interference
due to the lack of coordination,
the narrow beamwidth, increased channel loss,
and large bandwidth (hence large noise power) in the mmWave
scenario mean that noise and not interference is
usually the dominant effect \cite{mustafa-channel}.
In the microwave scenario, where intercell interference
is stronger, we use frequency reuse
as in~\cite{andrews2011tractable}.
For frequency reuse factor $\delta$,
each cell is randomly assigned one band with bandwidth
$\frac{W_i}{\delta}$ from the $\delta$ bands available
to each NSP.

Early work on resource sharing in mmWave
networks~\cite{andrews-sharing,european-sharing,matiamag,matia}
has focused on signal propagation
and interference effects in networks with shared resources.
To approximate the data rate at a UE, they divide
its link capacity as determined by SINR by the total number of UEs
in the cell. In a realistic network with opportunistic
scheduling, however, a UE may achieve a higher data rate
than its average SINR would suggest, because it is scheduled
with higher priority in time slots when its SINR is high.
For an economic analysis we need to accurately model
how consumers' utility scales with all aspects of
network size, including the number of subscribers,
so our model must include this scheduling gain.
We adopt a modified scheduler based on
the multicell temporal fair opportunistic scheduler proposed
in~\cite{shahram-scheduler}. That scheduler involves two stages:
in the first stage a UE is nominated for each cell,
and then, after some coordination among base stations,
a subset of the nominated users are scheduled in the second stage.
We have no coordination between BSs, so we use only the
first (user nomination) stage of the scheduler mentioned above.
Thus each BS runs the scheduler and selects a UE
independently, without considering intercell interference.

\section{Network simulation results}
\label{sec:simulation}

Using the model of Section~\ref{sec:system},
we simulate mmWave and microwave networks
with the parameters given in
Table~\ref{sim-param}.

\begin{table}[h]
\caption{Network parameters}
\centering
\begin{tabular}{lll}
\toprule \textbf{Parameter} & \textbf{mmWave}   & \textbf{microwave} \\
\midrule
Frequency                         &  73 GHz                       & 2.5 GHz \\
Max. bandwidth ($W_{\text{max}}$)        &  1 GHz                        & 300 MHz \\
Frequency reuse factor            &  1                            & 3 \\
Max. BS density ($\lambda^B_{\text{max}}$)     &  100 BSs/$\text{km}^2$        &  100 BSs/$\text{km}^2$  \\
Max. UE density ($\lambda^U_{\text{max}}$)     &  500 UEs/$\text{km}^2$        &  500 UEs/$\text{km}^2$  \\
BS transmit power $P$             &  30 dBm                       &  30 dBm \\
($M^B$,$m^B$,$\theta^B$)          & (20 dB, -10 dB, 5\degree)     & (0 dB, -20 dB, 70\degree) \\
($M^U$,$m^U$,$\theta^U$)          & (10 dB, -10 dB, 30\degree)    & (0 dB, 0 dB, 360\degree)\\
Simulation area                              & 1 $\text{km}^2$                & 1 $\text{km}^2$  \\
Rate model ($\alpha$, $\beta$)    & (0.2, 0.5)                    & (0.2, 0.5) \\
UE noise figure $N_f$             & 7 dB                          & 7 dB\\
Noise PSD $N_0$                   & -174 dBm/Hz                   & -174 dBm/Hz\\
\bottomrule
\end{tabular}
\label{sim-param}
\end{table}

Recall that we model mmWave network service
as a \emph{network good}, as described in Section~\ref{sec:economic}.
Subscribers benefit from an
indirect positive network externality: a
large NSP with more
subscribers (higher density of UEs) will build a denser
deployment of BSs and purchase more spectrum.
Thus network size of NSP $i$, $n_i$, represents
the normalized demand for the service
(scaled to the range $[0,1]$),
but is also a scaling factor on the BS
density and bandwidth of the NSP. Specifically, as $n_i$ varies,
the UE density $\lambda_i^{U}$ is  $ n_i \lambda_{\text{max}}^{U}$,
the BS density $\lambda_i^{B}$ is  $ n_i \lambda_{\text{max}}^{B}$,
and the bandwidth $W_i$ is $n_i W_{\text{max}}$.
We then find the net effect of increasing network size $n_i$
on subscribers' fifth percentile data rates.
We use fifth percentile rates as a key metric of utility
because research on human behavior
suggests that service reliability is rated highly
in perceived quality of mobile service~\cite{Kuo2009887}.
We take the fifth percentile
rate as a proxy for service reliability to obtain $h(n)$, 
the network externalities function introduced
in Section~\ref{sec:economic}.

\begin{figure}[t!]
    \centering
    \begin{subfigure}[t]{0.47\linewidth}
        \centering
        \includegraphics[width=\textwidth]{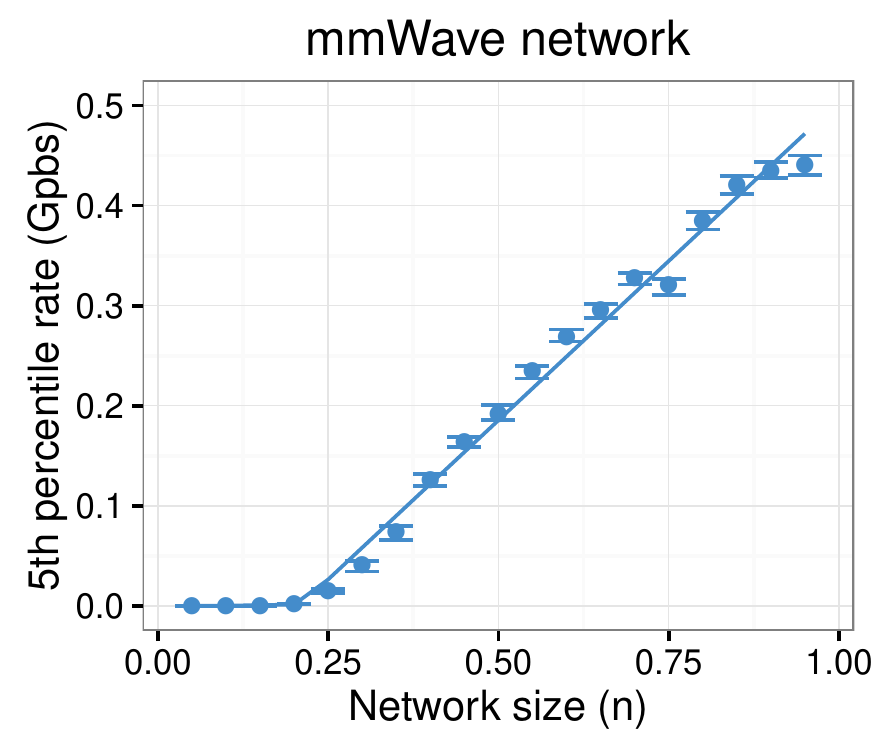}
        \caption{}
        \label{fig:netmmwave}
    \end{subfigure}%
    \begin{subfigure}[t]{0.47\linewidth}
        \centering
        \includegraphics[width=\textwidth]{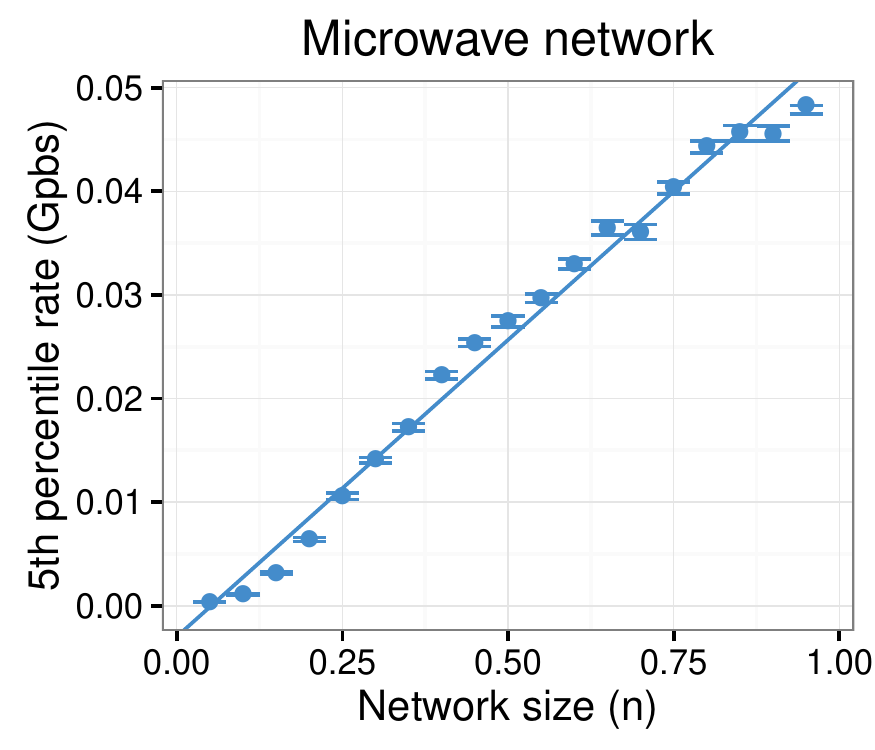}
        \caption{}
        \label{fig:netmicrowave}
    \end{subfigure}
    \caption{Effect of increasing network size on fifth percentile
    data rates (used as a metric of consumer utility), with the
    piecewise line estimated by an ordinary least squares linear regression.
    Error bars show 95\% bootstrap confidence intervals. }
     \label{fig:neteffects}
\end{figure}

The simulation results are shown in Fig.~\ref{fig:neteffects}.
We approximate the variation of the fifth percentile rate
with network size as a
linear or piecewise linear relationship,
estimated by an ordinary least squares linear regression,
with breakpoints found using the \texttt{segmented}
package for R~\cite{muggeo2003estimating}.

We note two key differences between mmWave
and microwave networks.
First, in microwave networks,
the fifth percentile rate increases
beginning from a very small network size.
In mmWave networks,
the fifth percentile rate remains flat at first
and starts increasing only at a moderate network size.
This is due to the increased path loss
at mmWave frequencies, where a denser
deployment of BSs is necessary to prevent outage.
Second, we note that for moderate or large networks,
the network effect is stronger
in mmWave networks, as per the
slope of the line in Fig.~\ref{fig:netmmwave}
relative to the line in Fig.~\ref{fig:netmicrowave} (note the
different vertical axis scales).
This is due to the much larger bandwidth and
the greater benefit due to having a LOS
link in a mmWave network.

Fig.~\ref{fig:neteffects} also suggests some intuition
regarding incentives for sharing in mmWave
and microwave networks.
The effect of resource sharing is similar to
an increase in network size:
when two NSPs of size $n_1$
and $n_2$ agree to share resources, subscribers of both NSPs experience
the network effects due to the total network size $n_1 + n_2$.
Fig.~\ref{fig:neteffects} shows
that network effect in mmWave networks is different,
and that the network
size has a greater impact on utility than in microwave networks.
Thus resource sharing may be
less profitable for a dominant (large $n$) mmWave service provider,
because the advantage gained by its competitor due
to sharing is greater than it would be in an equivalent
microwave network scenario.
In the next section, we will examine this intuition
in the context of a duopoly game.

\section{Duopoly game}
\label{sec:duopoly}

We model the NSPs' decision to share network resources
or not as a compatibility problem (introduced in Section~\ref{sec:economic}),
where NSPs are considered
compatible if their subscribers can connect to
any of the set of NSPs' BSs, and use
a bandwidth equal to their pooled spectrum holdings.
In many locations in the United States and around the world,
the market for cellular service is effectively a
duopoly. We therefore consider a market
with two vertically differentiated NSPs,
and the three-stage game with complete information
described in~\cite{baake2001vertical}:

\begin{enumerate}
\item NSPs $i\in\{1,2\}$ simultaneously choose quality $q_i$ from $[0, \hat{q}]$.
\item NSPs $i\in\{1,2\}$ simultaneously set price $p_i$.
\item Each consumer subscribes to one NSP $i\in\{1,2\}$
or neither.
\end{enumerate}

The quality $q_i$ is the \emph{inherent} quality (with maximum feasible value $\hat{q}$).
It refers to aspects of service unrelated to the
size of the network, such as
the quality of the legacy data network,
customer service, and the availability of desirable handsets.

An NSP's marginal costs are increasing in $q_i$,
with cost function
$c(q_i, n_i) = q_i n_i$,
and so each NSP $i \in \{1,2\}$ seeks to maximize its profits
\begin{equation}
\pi_i(q_i, n_i, p_i) = n_i p_i - q_i n_i
\label{eq:nspprofit}
\end{equation}

Consumers evaluate competing services in terms of the difference in
their inherent qualities as well as their network externalities.
We have heterogeneous consumers parameterized by $\omega$,
with $\omega$ distributed uniformly from $[0, \hat{\omega}]$.
The surplus of a consumer of type $\omega$ subscribing 
to NSP $i$ is given by
\begin{equation}
 u(\omega, q_i, \tilde{n}_i, p_i ) =
   \omega q_i + q_i h(\tilde{n}) - p_i 
\end{equation}
\noindent with $i \in \{1,2\}$,
and each consumer subscribes to at most one NSP.
If the NSPs share their mmWave network resources,
then $\tilde{n}_i = \sum_{i \in \{1,2\}} n_{i}$, otherwise $\tilde{n}_i = n_i$.

The fifth percentile rate is
linear or piecewise linear
in the network size in a mmWave (Fig.~\ref{fig:netmmwave})
or microwave network (Fig.~\ref{fig:netmicrowave}).
We consider only moderate- to large-sized
networks, corresponding to the line on the right side
of Fig.~\ref{fig:netmmwave}.
Thus the network externalities
function $h(\tilde{n}_i)$ is approximately linear in $\tilde{n}_i$, 
and we take $h(\tilde{n}_i) = \mu \tilde{n}_i $
The scaling factor $\mu$ determines the intensity of the network
externality,
and is derived from slopes of the lines
in Fig.~\ref{fig:neteffects} as $\mu_{mm} = 0.64$
and $\mu_{\text{micro}} = 0.05$.

\begin{figure*}[t!]
    \centering
    \begin{subfigure}[t]{0.215\linewidth}
        \centering
        \includegraphics[width=\textwidth]{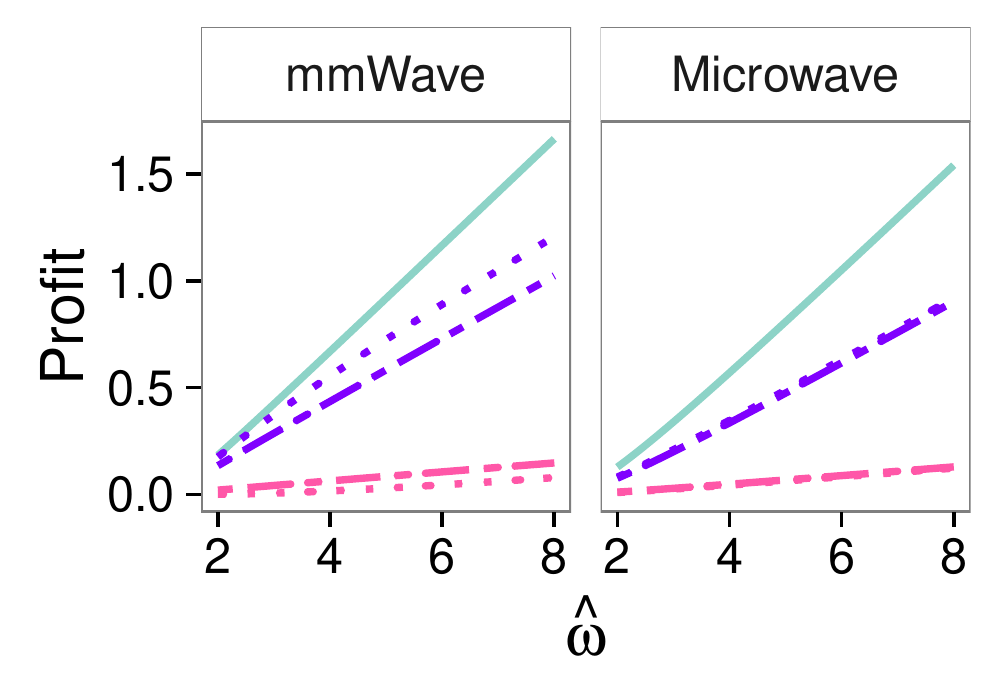}
        \caption{NSP profit, $\hat{q} = 1$}
        \label{fig:profit1}
    \end{subfigure}%
    \begin{subfigure}[t]{0.215\linewidth}
        \centering
        \includegraphics[width=\textwidth]{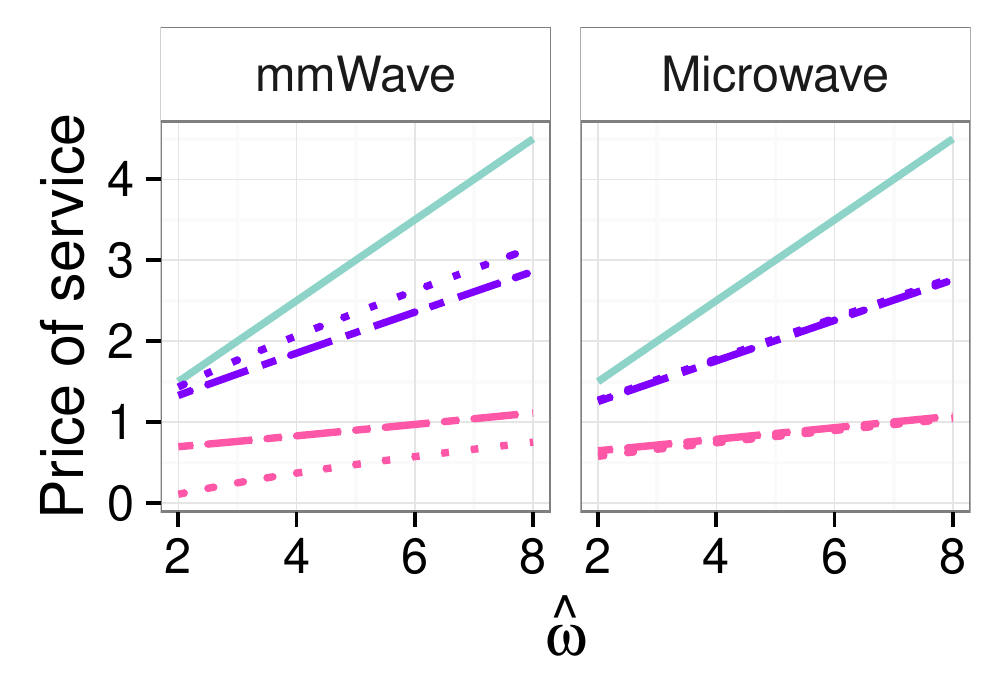}
        \caption{Price, $\hat{q} = 1$}
        \label{fig:price1}
    \end{subfigure}
    \begin{subfigure}[t]{0.215\linewidth}
        \centering
        \includegraphics[width=\textwidth]{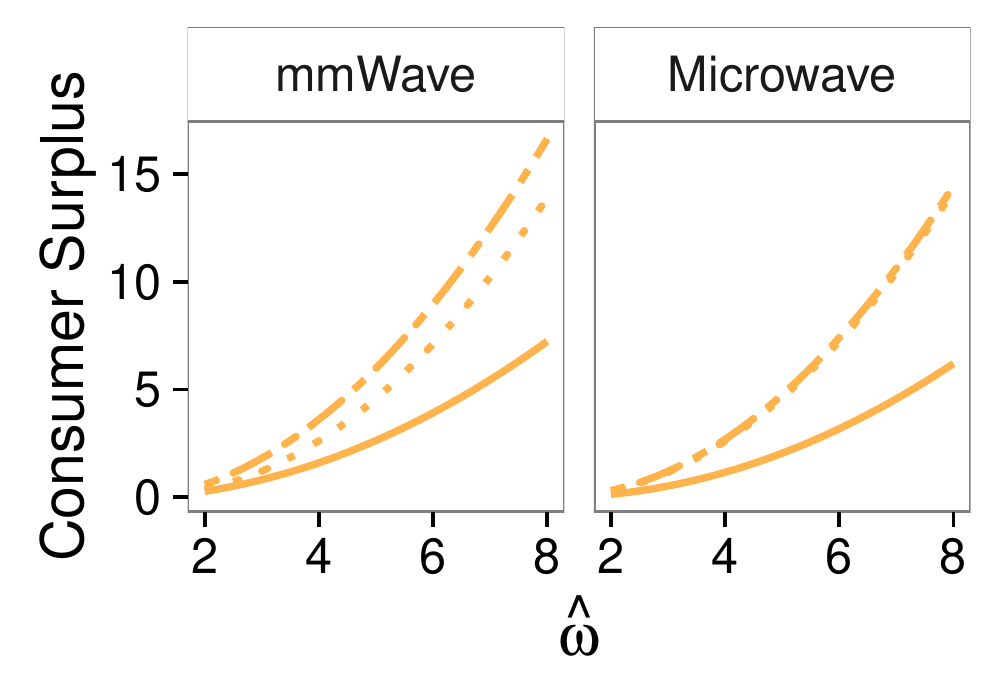}
        \caption{Consumer surplus, $\hat{q} = 1$}
        \label{fig:csurplus1}
    \end{subfigure}%
    \begin{subfigure}[t]{0.348\linewidth}
        \centering
        \includegraphics[width=\textwidth]{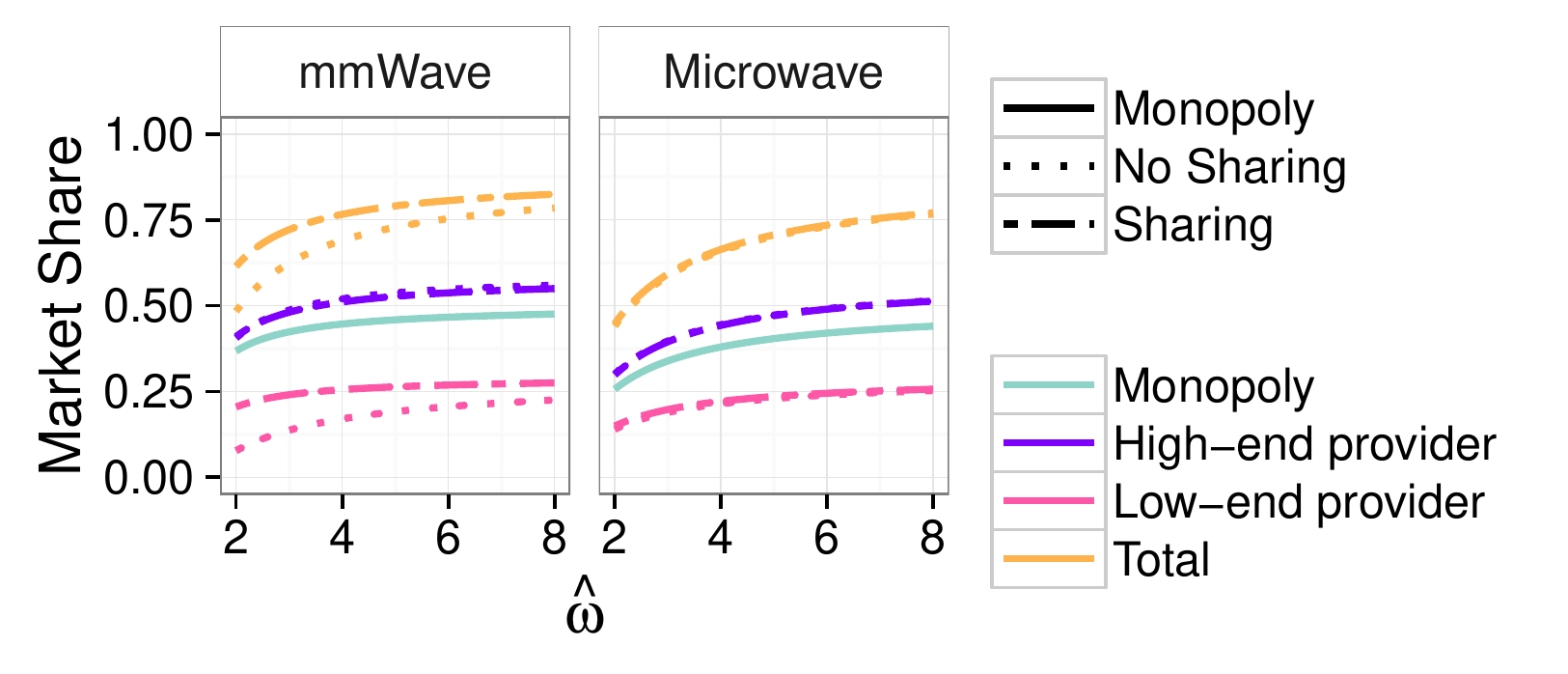}
        \caption{Market share, $\hat{q} = 1$}
        \label{fig:share1}
        \vspace*{3mm}
    \end{subfigure}
    \begin{subfigure}[t]{0.215\linewidth}
        \centering
        \includegraphics[width=\textwidth]{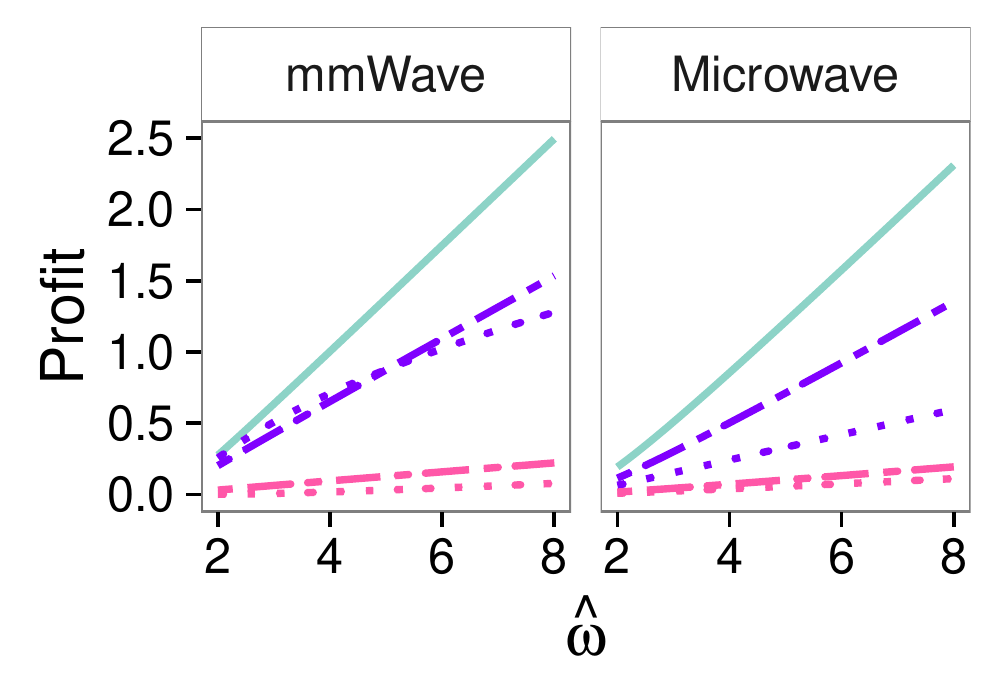}
        \caption{NSP profit, $\hat{q} = 1.5$}
        \label{fig:profit15}
    \end{subfigure}%
    \begin{subfigure}[t]{0.215\linewidth}
        \centering
        \includegraphics[width=\textwidth]{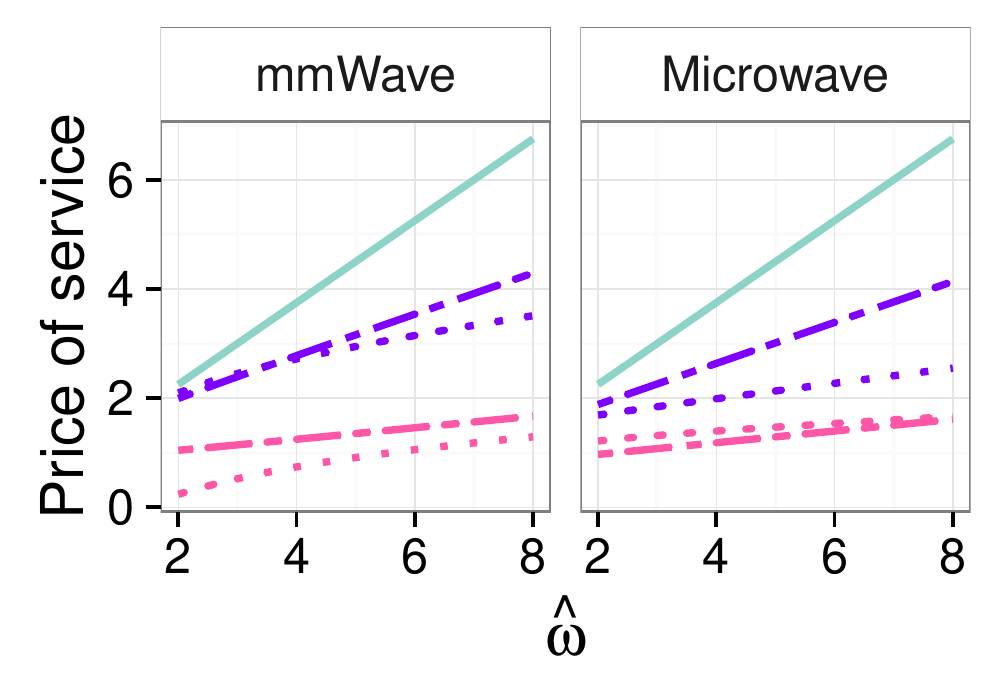}
        \caption{Price, $\hat{q} = 1.5$}
        \label{fig:price15}
    \end{subfigure}
    \begin{subfigure}[t]{0.215\linewidth}
        \centering
        \includegraphics[width=\textwidth]{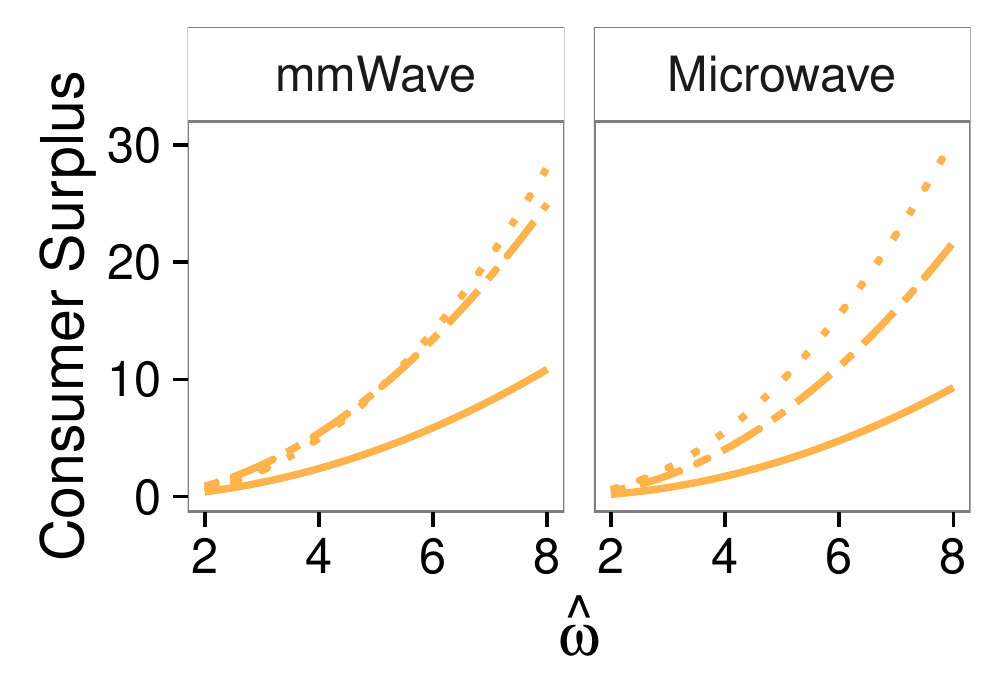}
        \caption{Consumer surplus, $\hat{q} = 1.5$}
        \label{fig:csurplus15}
    \end{subfigure}%
    \begin{subfigure}[t]{0.348\linewidth}
        \centering
        \includegraphics[width=\textwidth]{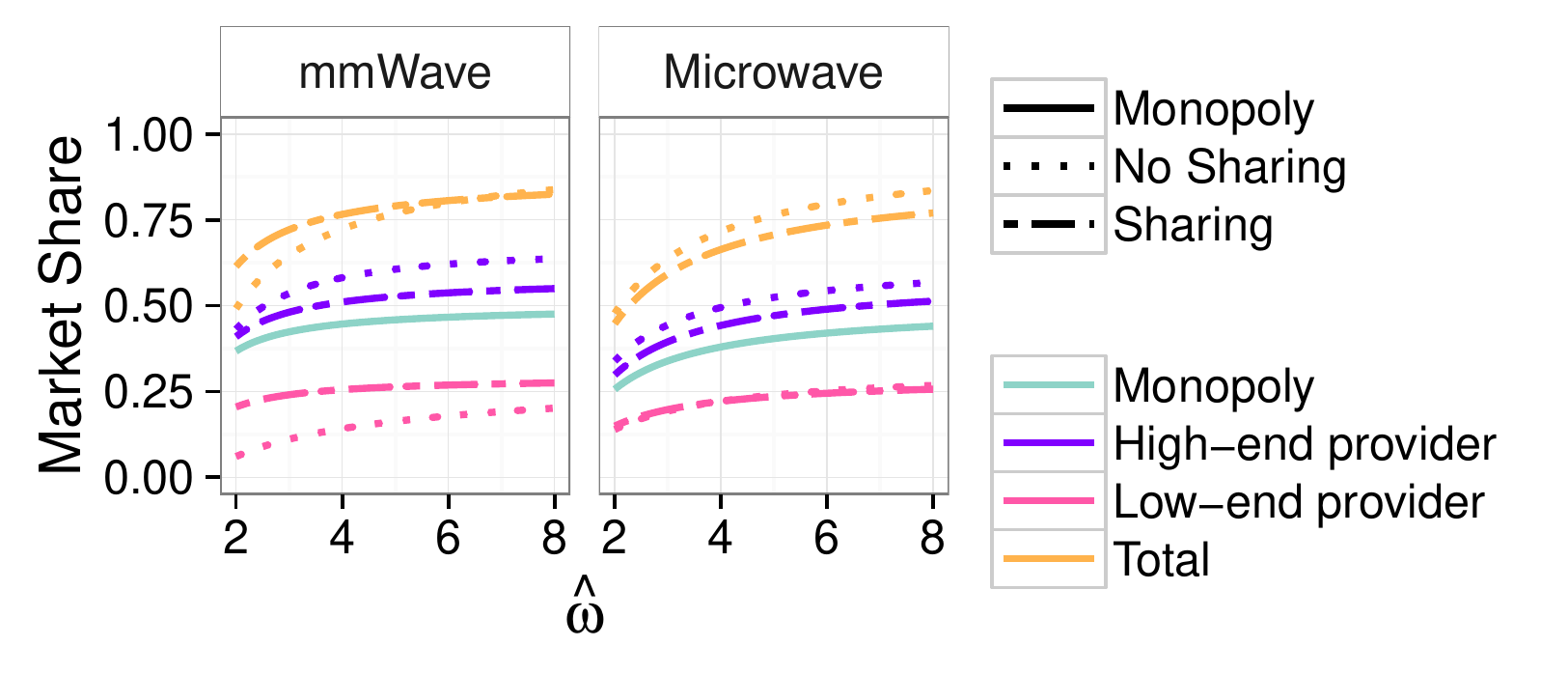}
        \caption{Market share, $\hat{q}=1.5$}
        \label{fig:share15}
    \end{subfigure}
    \caption{Profit of each NSP, price set by NSP, total consumer surplus, and market
    share of each NSP (and total market coverage) in the duopoly game. As $\hat{\omega}$ increases, the market is more easily separated into high-end
    and low-end groups and NSPs compete primarily on quality, so there is
    less price competition. A similar effect occurs as
    $\hat{q}$ increases. The intensity of the network effect is
    $\mu_{mm} = 0.64$ for mmWave networks and $\mu_{\text{micro}} = 0.05$
    for microwave networks.}
     \label{fig:duopoly}
\end{figure*}

By their choice of quality level, the NSPs
segment the market into a low-end group (small-$\omega$ type)
and a high-end group (large-$\omega$ type).
Without loss of generality, we assign index 1 to the NSP 
that chooses the higher quality, i.e., $q_1 > q_2$, and
subscribers of NSP 1 belong to the
large-$\omega$ group.
We define two marginal consumers: the consumer of type
$\underline{\omega}$ is indifferent between choosing no subscription
and subscribing to NSP 2,
and the consumer of type $\overline{\omega}$ is indifferent between
subscribing to NSP 1 and subscribing to NSP 2.

Then the utility of the marginal consumer of type $\overline{\omega}$
satisfies
\begin{equation}
\overline{\omega} q_1 + \mu q_1 \tilde{n}_1 - p_1  = \overline{\omega} q_2 + \mu q_2 \tilde{n}_2 - p_2
\label{omegaover}
\end{equation}
and the utility of the marginal consumer of type $\underline{\omega}$
satisfies
\begin{equation}
\underline{\omega} q_2 + \mu q_2 \tilde{n}_2 - p_2  = 0
\label{omegaunder}
\end{equation}

Also, the marginal consumer of type $\overline{\omega}$
defines the market share of the high-end service
\begin{equation}
n_1 = \frac{\hat{\omega}-\overline{\omega}}{\hat{\omega}}
\label{n1share}
\end{equation}
and the marginal consumers together define
the market share of the low-end service
\begin{equation}
n_2 = \frac{\overline{\omega}-\underline{\omega}}{\hat{\omega}}
\label{n2share}
\end{equation}

We can solve (\ref{omegaover})-(\ref{n2share})
for $n_1$, $n_2$, $\overline{\omega}$, and $\underline{\omega}$,
and thus determine the decisions of the consumers and the market
share of each NSP given $p_i, q_i, i \in \{1,2\}$.

It is
shown in \cite{baake2001vertical} that when 
$0 \leq \mu < \text{min}[1, \hat{\omega}/2]$
and the ratio of quality levels satisfies
\begin{equation}
\frac{q_1}{q_2} > \Bigg( \frac{\hat{\omega}^2}{(\hat{\omega}-\mu)(\hat{\omega}-2\mu)}\Bigg)
\label{highprice}
\end{equation}
there is a unique Nash equilibrium with both
NSPs' prices higher than their marginal costs.
Furthermore, if
(\ref{omegaover})-(\ref{n2share})
satisfies
\begin{equation}
0 < \underline{\omega} < \overline{\omega} < \hat{\omega}
\label{marketshare}
\end{equation}
then both NSPs have market share greater than zero.
When both (\ref{highprice}) and (\ref{marketshare}) hold, then there is a unique
Nash equilibrium in which both NSPs earn non-zero profit.
We are primarily interested in the scenario 
in which both service providers offer service at a non-zero profit, 
so we restrict our attention to these circumstances.

If (\ref{highprice}) and (\ref{marketshare}) hold and
NSPs do not share resources,
then per \cite{baake2001vertical} their equilibrium
quality levels $q_{1,NS}^*, q_{2,NS}^*$ are:
\begin{equation}
q_{1,NS}^* = \hat{q}
\end{equation}
\makeatletter
\if@twocolumn%
    \begin{equation}
    \resizebox{0.42\textwidth}{!}{
    $q_{2,NS}^* = \frac{\hat{q}(\hat{\omega}-\mu)^2 \Big[11\hat{\omega}-10\mu - \sqrt{3(3\hat{\omega}^2 + 28\hat{\omega}\mu - 20\mu^2)}\Big]}{2\hat{\omega}^2(7\hat{\omega}-5\mu)}  < \hat{q}$
    }
    \end{equation}
\else
    \begin{equation}
    q_{2,NS}^* = \frac{\hat{q}(\hat{\omega}-\mu)^2 \Big[11\hat{\omega}-10\mu - \sqrt{3(3\hat{\omega}^2 + 28\hat{\omega}\mu - 20\mu^2)}\Big]}{2\hat{\omega}^2(7\hat{\omega}-5\mu)}  < \hat{q}
    \end{equation}
\fi
\makeatother

\noindent and their equilibrium prices 
$p_{1,NS}^*, p_{2,NS}^*$ are:
    \begin{equation}
        \resizebox{0.42\textwidth}{!}{
$p_{1,NS}^* = q_1 \Bigg[1 + \frac{(\hat{\omega}-1)[2q_1(\hat{\omega}-\mu)^2 - q_2 \hat{\omega}(2\hat{\omega}-\mu)]}{4q_1(\hat{\omega}-\mu)^2 - q_2\hat{\omega}^2} \Bigg] > q_1$
}
\label{eq:price1nosharing}
\end{equation}
    \begin{equation}
        \resizebox{0.42\textwidth}{!}{
$p_{2,NS}^* = q_2 \Bigg[1 + \frac{(\hat{\omega}-1)[q_1(\hat{\omega}-\mu)(\hat{\omega}-2\mu) - q_2\hat{\omega}^2]}{4q_1(\hat{\omega}-\mu)^2 - q_2\hat{\omega}^2} \Bigg] > q_2$
}
\label{eq:price2nosharing}
\end{equation}
\noindent The profits of the high-end NSP always increase with $q_1$,
so it will use $\hat{q}$. The low-end NSP balances
two competing effects: at high values of $q_2$ it
captures more of the market, but is also more similar to $q_1$, which
increases price competition.

The total consumer surplus is then
\begin{equation}
\resizebox{0.42\textwidth}{!}{
$\int_{\overline{\omega}}^{\hat{\omega}} u(\omega, q_{1,NS}^*, n_1, p_{1,NS}^* ) \; d\omega + 
\int_{\underline{\omega}}^{\overline{\omega}} u(\omega, q_{2,NS}^*, n_2, p_{2,NS}^* ) \; d\omega $
}
\label{eq:csurplusnosharing}
\end{equation}

If the NSPs share resources,
then per \cite{baake2001vertical} 
their equilibrium
quality levels $q_{1,S}^{*}, q_{2,S}^{*}$ are:
\begin{equation}{}
q_{1,S}^{*} = \hat{q}
\end{equation}
\begin{equation}
q_{2,S}^{*} =  \frac{\hat{q}(4\hat{\omega}-3\mu)}{7\hat{\omega}-6\mu} < \hat{q}
\end{equation}

\noindent and their equilibrium
prices $p_{1, S}^{*}, p_{2, S}^{*}$ are:
\begin{equation}
p_{1, S}^{*} = q_1 \Bigg[ 1 + \frac{2\hat{\omega}(\hat{\omega}-1)(q_1-q_2)}{(4\hat{\omega}-3\mu)q_1 - \hat{\omega}q_2} \Bigg] > q_1
\label{eq:price1sharing}
\end{equation}
\begin{equation}
p_{2, S}^{*} = q_2 \Bigg[ 1 + \frac{\hat{\omega}(\hat{\omega}-1)(q_1-q_2)}{(4\hat{\omega}-3\mu)q_1 - \hat{\omega}q_2}  \Bigg] > q_2
\label{eq:price2sharing}
\end{equation}
The total consumer surplus is then
\begin{equation}
\resizebox{0.42\textwidth}{!}{
$\int_{\overline{\omega}}^{\hat{\omega}} u(\omega, q_{1,S}^*, n_1 + n_2, p_{1,S}^* ) \; d\omega + 
\int_{\underline{\omega}}^{\overline{\omega}} u(\omega, q_{2,S}^*, n_1 + n_2 , p_{2,S}^* ) \; d\omega $
}
\label{eq:csurplussharing}
\end{equation}

We are also interested in the
monopoly case, for comparison. When there is one NSP,
the marginal consumer is defined by
\begin{equation}
\overline{\omega} q_1 + \mu q_1 \tilde{n}_1 - p_1  = 0
\end{equation}

\noindent and the market share of the NSP is
\begin{equation}
n_1 = \frac{\hat{\omega}-\overline{\omega}}{\hat{\omega}}
\label{eq:sharemonopoly}
\end{equation}
At equilibrium, the monopoly NSP will choose 
quality level
\begin{equation}
q_{1,M}^{*} = \hat{q}
\end{equation}
and price
\begin{equation}
p_{1,M}^{*} = \frac{q_1(\hat{\omega}-1)}{2}
\label{eq:pricemonopoly}
\end{equation}
The total consumer surplus is then
\begin{equation}
\int_{\overline{\omega}}^{\hat{\omega}} u(\omega, q_{1,M}^*, n_1 , p_{1,M}^* ) \; d\omega
\label{eq:csurplusmonopoly}
\end{equation}

Fig.~\ref{fig:duopoly} shows the
NSP profit (\ref{eq:nspprofit}); price 
(\ref{eq:price1nosharing}), (\ref{eq:price2nosharing}), 
(\ref{eq:price1sharing}), (\ref{eq:price2sharing}), and (\ref{eq:pricemonopoly}); 
total consumer surplus (\ref{eq:csurplusnosharing}), (\ref{eq:csurplussharing}), 
and (\ref{eq:csurplusmonopoly});
and market share (\ref{n1share}), (\ref{n2share}), and (\ref{eq:sharemonopoly})
in the duopoly game, for various market
parameters ($\hat{\omega}$, $\hat{q}$). 
First, we note that our intuition
of Section~\ref{sec:simulation} is validated:
Fig.~\ref{fig:profit1}
and Fig.~\ref{fig:profit15} show that resource sharing is
less often profitable to both NSPs at the same time in mmWave networks
than in microwave networks. 
The low-end NSP benefits more from sharing in 
mmWave networks because of the greater importance 
of having a large network.
For the same reason, however, the competitive advantage
that comes with \emph{not} sharing for the dominant (high-end) NSP
is greater than in microwave networks. We note from 
Fig.~\ref{fig:price1} and Fig.~\ref{fig:price15} that 
under market conditions where the high-end NSP does not prefer sharing, 
it sets lower prices in the sharing scenario than the no sharing 
scenario, indicating increased price competition 
from the low-end NSP when sharing resources. 
We also note that with sharing, some previously high-end 
subscribers will subscribe to the low-end NSP
(Fig.~\ref{fig:share1} and Fig.~\ref{fig:share15}).

We further find that the consumer does not necessarily
enjoy greater surplus when NSPs share resources
(Fig.~\ref{fig:csurplus1} and Fig.~\ref{fig:csurplus15}).
This is because an NSP may raise prices
(Fig.~\ref{fig:price1} and Fig.~\ref{fig:price15}), especially
when price competition is not a concern.
In particular, the low-end NSP in mmWave networks  
is likely to raise its price because its service
is dramatically improved by sharing.

The benefit to NSPs and consumers of resource sharing
depends on the market parameters, $\hat{q}$ and $\hat{\omega}$.
Reducing $\hat{q}$, the maximum possible quality level, 
increases price competition;
when the difference in quality levels between NSPs
is small, consumers are more sensitive to price.
Similarly, reducing the maximum value of the taste parameter, $\hat{\omega}$,
increases price competition, since this decreases the dispersion
of consumers' willingness to pay and the market is less segmented.
These effects are magnified when the intensity of the network
effect is large (as it is in mmWave networks); then,
the high-end NSP prefers resource sharing only
when the market is highly segmented and there is little price competition
(i.e., for large $\hat{\omega}$ and $\hat{q}$).

\section{Conclusions}
\label{sec:conclusion}

In this work, we have modeled the resource sharing decisions of
service providers in mmWave 5G cellular networks.
We have described a duopoly game involving two vertically differentiated
cellular service providers,
and compared mmWave and microwave networks
with respect to key economic metrics.

Our results suggest that resource sharing is profitable
for both NSPs at the same time less often in mmWave networks than
in microwave networks. Because resource sharing
has a stronger impact on subscribers'
data rate in mmWave networks, the competitive
advantage held by the larger NSP due to \emph{not} sharing resources
is greater and more likely to offset
the gains associated with consumers' increased
willingness to pay for service in a network
with resource sharing.  We also find that because
NSPs may increase prices when they share resources
(because of subscribers' greater willingness to pay
for a large network), resource sharing
does not necessarily have a net positive effect on consumer
surplus,
thus regulation that mandates resource sharing
may not always be in consumers' best interests.
However, regulators may still consider mandated 
sharing under circumstances where, due to 
factors not captured in this model, 
the low-end NSP would choose to leave the market.
In this case, mandated sharing increases the low-end NSP's
profits and might encourage it to stay in the market, 
improving consumer surplus relative to a monopoly.

We briefly discuss here some limitations 
of our approach.
Our model captures key factors in consumer
decisions, including the price of service, 
the size of the spectrum and base station resources
available to subscribers (incorporating both 
the service provider's resources and shared resources), 
and factors affecting the perceived value of service
that are not related to the network size.
Similarly, from the service providers' perspectives, 
our model includes the effects 
of price competition between service providers, 
the dispersion in consumers' willingness to pay
for service, and the increased cost of offering 
a service with a greater inherent quality.
However, our model does not directly include the 
initial investment cost associated with deploying a mmWave cellular
network, such as spectrum licensing costs; 
because the licensing mechanism for mmWave bands
has not yet been decided, it is difficult 
to model accurately at this time. 
Similarly, because mmWave cellular service has not yet been 
deployed in any market, we cannot use empirical data
from the mmWave cellular service market to inform or validate our analysis.
Our model uses the utility function whose form is proposed in~\cite{baake2001vertical},
in which the inherent quality (vertical differentiation factor) 
multiplies the quality factor that is related 
to network size. As a result, it may overestimate the magnitude of 
the disparity between service providers' profits and prices. 
However, other models 
of duopoly markets with network effects without vertical differentiation,
such as~\cite{economides1997compatibility},
similarly find that compatibility is not a consensual equilibrium 
when the magnitude of the network effect is large.

As future work, we would like to explore alternative
approaches to sharing, to preserve the technical
sharing gains while also improving NSP profit and
consumer surplus.

\section*{Acknowledgments}

This work was supported by the National Science Foundation
under Grant No. 1547332, 1302336, and the GRFP, by
the New York State Center for Advanced Technology in
Telecommunications (CATT),
and by NYU WIRELESS.

\ifCLASSOPTIONcaptionsoff
  \newpage
\fi

\bibliographystyle{IEEEtran}
\bibliography{jsac}

\end{document}